\documentclass[10pt,a4paper]{article}

\usepackage[latin1]{inputenc}
\usepackage{amsmath}
\usepackage{amsfonts}
\usepackage{amssymb}
\usepackage[pdftex]{graphicx}
\usepackage{multicol}
\usepackage[usenames]{color}
\usepackage{url}

\begin{document}
%
\title{A Multipath Transport Protocol for Future Internet}

\author{Bachir Chihani - Denis Collange\\
Orange Labs  Sophia Antipolis, France\\
Email: \{bachir.chihani,denis.collange\}@orange-ftgroup.com\\
}
\maketitle

\section{Aims}
Our work aims to assess a multipath transport protocol named MPTCP (Multipath TCP), especially the various congestion control and coupling mechanisms considered for this protocol.
Many authors propose to take advantage of the multiple paths that are often available for a data flow to improve its performance and its robustness to the varying transmission conditions. The different layers of the TCP/IP stack have been considered. At the application level, some authors suggest to handle many simultaneous TCP sockets and to shuffle the data upon them, like PSockets (Parallel Sockets). Other authors propose to open many end-to-end sub-connections in the transport layer transparently for the application, like cTCP (Concurrent TCP). At the network layer, some routing protocols may use multiple paths to route packets, like CMR (Concurrent Multipath Routing). And some algorithms have also been proposed at the Link layer, to split the data flow over several channels, like McMAC (Multiple channel MAC).

Multiple paths between two end-points may have different characteristics (i.e. capacity, latency, etc.), or diverse and varying traffic conditions. In consequence, packets using different paths may arrive out of sequence to the receiver. In this case, according to the classical cumulative acknowledgement mechanism of TCP, the data receiver sends back duplicate acknowledgements. With the TCP mechanism Fast Retransmit the data source may then misinterpret these duplicate acknowledgements as an indication of packet loss due to a congestion in the network, and then wrongly reduce its throughput. 
An efficient multipath protocol has then to be able to deal with these diversity of characteristics to optimize the throughput and to react nonetheless properly to any change. 

Transport protocols can take advantage of feedback mechanisms, through the acknowledgements for example, to gather end-to-end up-to-date information about each path. This information can then be used to optimize dynamically the control of the traffic on the multiple available paths,  for instance moving the traffic away from a congested path. An IETF working group, Multipath TCP (MPTCP), has started in 2009 to specify a multipath transport protocol. The aims are to improve the performance of a multi-path flow, i.e. to be at least as good as a single-path flow on the best route, without consuming on any path more capacity than a single-path flow. A last objective is to balance the congestion, moving away the traffic from congested paths.  
With MPTCP protocol, a congestion control is performed on each path at the subflow level, and coupling mechanisms can be introduced to satisfy the design objectives. Many coupling methods have been proposed and compared, especially "Fully Coupled", "Linked Increases" and "RTT Compensator".

\section{Methods}
To analyse MPTCP performance and the way it reacts to different network situations, we have implemented the protocol as defined in the current MPTCP drafts (using their release of July 2010) under the NS-3 network simulator. We have also implemented the different methods proposed to couple the congestion control mechanisms between the multiple subflows. These diverse coupling methods mainly differ in the increase rate of their congestion window after an acknowledgement and in their window decrease after a packet loss.

Spurious loss detections due to the variety of path characteristics and traffic conditions remain an obstacle to achieve the optimal performance. We suggest to enhance MPTCP by adding packet reordering algorithms to face this problem. Many such mechanisms have been proposed to optimize the performance of Standard TCP in case of large jitter, for example on wireless access networks. These proposals may be grouped in two families: state reconciliation and delayed response. With mechanisms from the first family, the TCP source first enters the Fast Retransmit state. Then, when it detects the spurious lost, it reacts by returning to the state preceding the Fast Retransmit. With mechanisms from the second family, the sender avoids spurious retransmissions by delaying the response to congestion. 
We implement two packet reordering algorithms, from the first family: DSACK (Duplicate Selective ACKnowledgments) and Eifel. We choose these two as representative of their family because they are specified in IETF Requests For Comments (RFC) and they are already implemented in many operating systems. To detect spurious retransmissions, DSACK (RFC 2883) uses the SACK fields in the TCP header to declare the packets received correctly. Eifel (RFC 4015) uses the Timestamp TCP fields to distinguish the initial packet sent and from its eventual retransmission. 
We have adapted these two algorithms to the case of multipath and we have included them in our MPTCP implementation in order to assess their behaviour on NS-3 simulations.

\begin{figure}[ht]
\begin{center}
$\begin{array}{ccc}
\includegraphics[scale=0.14]{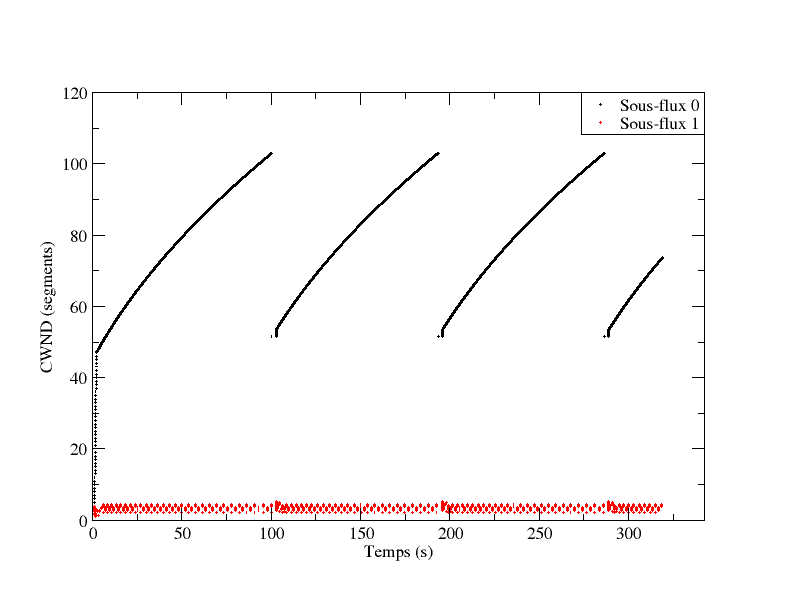} 
& \includegraphics[scale=0.14]{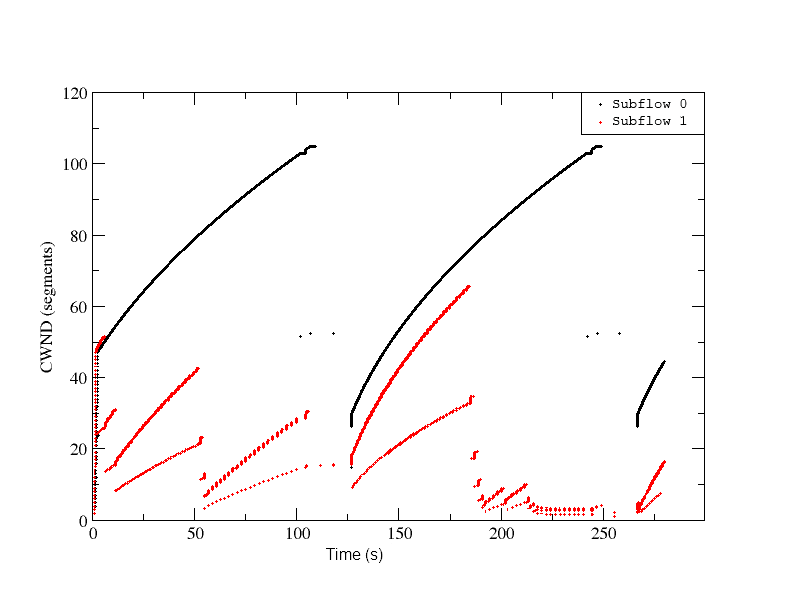} 
& \includegraphics[scale=0.14]{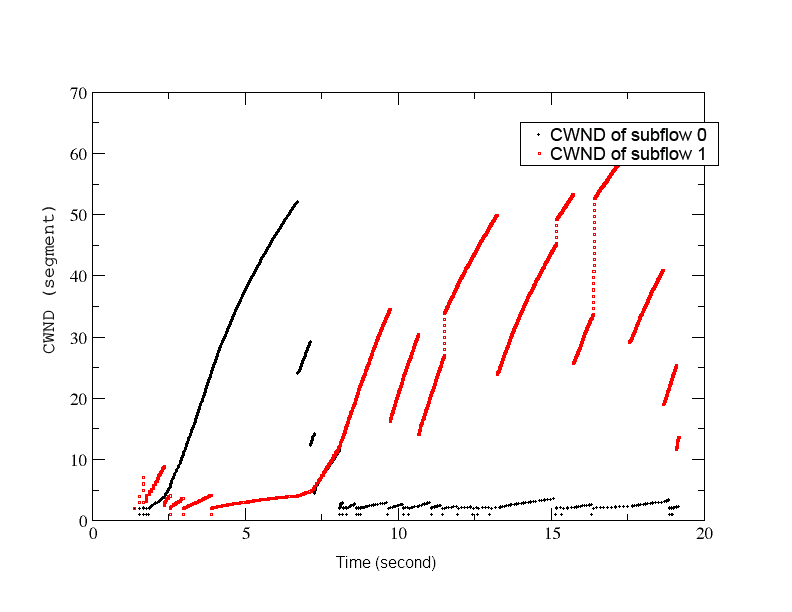} \\
\mbox{\bf (a)} & \mbox{\bf (b)} & \mbox{\bf (c)} \\
\end{array}$
\caption{Congestion windows progress in case of: (a) without a packet reordering algorithm, (b) Eifel algorithm, (c) D-SACK algorithm}
\label{cwnd}
\end{center}
\end{figure}

\section{Result}
We simulate a MPCTP connection with two available paths to transfer a 10 GB file between two hosts interconnected through two point to point links. The characteristics of one link are fixed (capacity = 0.5 Mbps, latency = 10 ms and loss rate = 0\%) and we varied separately several characteristics of the second link: the capacity from 0.5 to 16 Mbps, the latency from 0 to 320 ms and the loss rate from 0 to 10\%. 

In Figure 1, we present the respective evolution of the congestion windows of each subflow in case (a) where no packet reordering algorithm is used, (b) where the Eifel algorithm is used, and (c) where the D-SACK is used. 
In the first case (a), we can see that one of the subflow is dominating and the data are only sent through it. 
In the second case (b), the congestion window of the subflow 1 oscillates between two curves of evolution: the congestion window takes its values according to the lower one in case of segment retransmission, and returns to the upper one when the Eifel algorithm detects that the retransmission was spurious. 
In the third case, we observe three periods of time during which the congestion window of subflow 1 grows unusually in an exponential way. These periods reflect the DSACK Slow Start which is triggered by the DSACK algorithm after a spurious retransmission.

The Eifel's reaction to a spurious retransmission detection leads to the injection of a burst of traffic in the network. DSACK injects more smoothly packets but it takes then more time to reach the performances preceding the retransmission. So these algorithms improve the overall performance of MPTCP, but they do not alleviate performance problems due to persistent and substantial segment reordering. New reordering mechanisms suiting better the specific problem of MPTCP should then be considered

\section{Conclusion}
In this paper, we present MPTCP as an interesting multipath transport protocol, we discussed our implementation of this protocol under the NS-3 simulator and the simulated scenarios for evaluating its performance. We also analyse the influence of two packet reordering algorithms, Eifel and DSACK,  on the congestion window behaviour, as they are standardized and already implemented in current operating systems to optimize the performance of standard TCP on wireless networks. 
We think that packet reordering mechanisms can highly improve the performance of a multipath protocol. However the ones we have considered do not fit optimally the multipath context. They help to detect spurious retransmission and push the connection into a state recovery. We think that it may be interesting to evaluate other mechanisms who are able to avoid spurious retransmission, for example by suspending the congestion response momentarily or adjust the slow start threshold and the retransmission time out.
In the future, we aim to use a real implementation of MPTCP and other packet reordering mechanisms to experiment their behaviours in a real environment. 

\end{document}